\documentclass{article}
\usepackage{microtype}
\usepackage{graphicx}
\usepackage{subfigure}
\usepackage{booktabs} 
\usepackage{hyperref}

\usepackage[accepted]{icml2024}
\usepackage{amsmath}
\usepackage{amssymb}
\usepackage{mathtools}
\usepackage{amsthm}
\usepackage{aas}
\usepackage[capitalize,noabbrev]{cleveref}
\theoremstyle{plain}

\theoremstyle{definition}

\theoremstyle{remark}

\usepackage[textsize=tiny]{todonotes}
\icmltitlerunning{Variable Star Light Curves in Koopman Space}
\begin{document}
\twocolumn[
\icmltitle{Variable Star Light Curves in Koopman Space}
\icmlsetsymbol{equal}{*}
\begin{icmlauthorlist}
\icmlauthor{Nicolas Mekhaël}{equal,B,C,G}
\icmlauthor{Mario Pasquato}{equal,A,B,C,D}
\icmlauthor{Gaia Carenini}{equal,E,F}
\icmlauthor{Vittorio F. Braga}{I,H}
\icmlauthor{Piero Trevisan}{I}
\icmlauthor{Giuseppe Bono}{L}
\icmlauthor{Yashar Hezaveh}{B,C,D,M}
\end{icmlauthorlist}
\icmlaffiliation{A}{INAF IASF-Milano, Milan, Italy}
\icmlaffiliation{B}{Dèpartment de Physique, Universitè de Montrèal, Montrèal, Canada}
\icmlaffiliation{C}{Mila -- Quebec Artificial Intelligence Institute, Montrèal, Canada}
\icmlaffiliation{D}{Ciela -- Computation and Astrophysical Data Analysis Institute, Montrèal, Canada}
\icmlaffiliation{E}{Départment d'Informatique, Ecole Normale Supérieure -- Université Paris Sciences \& Lettres (PSL), Paris, France}
\icmlaffiliation{F}{Department of Computer Science and Technology, University of Cambridge, Cambridge, UK}
\icmlaffiliation{G}{Dèpartment de Physique, Université de Sherbrooke, Sherbrooke, Canada }
\icmlaffiliation{H}{INAF-Osservatorio Astronomico di Roma, Monte Porzio Catone, Italy}
\icmlaffiliation{I}{Space Science Data Center, Rome, Italy}
\icmlaffiliation{L}{Departimento di Fisica, Università di Roma Tor Vergata, Rome, Italy}
\icmlaffiliation{M}{Center for Computational Astrophysics, Flatiron Institute, New York, USA}
\icmlcorrespondingauthor{Mario Pasquato}{mario.pasquato@umontreal.ca}
\icmlcorrespondingauthor{Gaia Carenini}{gc645@cam.ac.uk}

\icmlkeywords{Machine Learning, ICML}

\vskip 0.3in
]

\printAffiliationsAndNotice{\icmlEqualContribution} 

\begin{abstract}
We present the first application of data-driven techniques for dynamical system analysis based on Koopman theory to variable stars. We focus on light curves of RRLyrae type variables, in the Galactic globular cluster $\omega$ Centauri. Light curves are thus summarized by a handful of complex eigenvalues, corresponding to oscillatory or fading dynamical modes. We find that variable stars of the RRc subclass can be summarized in terms of fewer ($\approx 8$) eigenvalues, while RRab need comparatively more ($\approx 12$). This result can be leveraged for classification and reflects the simpler structure of RRc light curves. We then consider variable stars displaying secular variations due to the Tseraskaya–Blazhko effect and find a change in relevant eigenvalues with time, with possible implications for the physical interpretation of the effect.
\end{abstract}
\section{Introduction}
In the context of time-domain astronomy and stellar variability, machine learning techniques are growing in popularity, particularly in light of the substantial amounts of data generated by current \citep[][]{2004SPIE.5489...11K, 2010Sci...327..977B, drake2013a, drake2014, 2016A&A...595A...2G, 2018MNRAS.477.3145J, 2021AJ....161..267V, 2021MNRAS.505.2954C} and planned ground- and space-based observing facilities. In the study of variable stars, these techniques have primarily been applied directly to light curves, which are represented by appropriately defined features (e.g. see \citealt{2018MNRAS.475.2326P, 2019PASP..131c8002M, 2021ApJ...920...33D, 2022MNRAS.514.2793B} for optical light curves, but also e.g. \citealt{2022A&A...659A..66K} for X-ray variable sources); deep neural networks have also been employed \citep[e.g.][]{2020ApJS..250...30J, 2021FrASS...8..168B}. Although these methods produce good predictions, they are typically not interpretable nor do they necessarily learn a correct representation of the underlying physics.

Here, we are interested in inherently interpretable techniques that, even though they remain fully data driven, allow us to get closer to the physics of the dynamical system producing the data. An example along these lines can be found in \cite{2022ApJ...930..161P}, where the authors use the sparse identification of nonlinear dynamics (SINDy) method to automatically identify governing equations for RRab, RRc, and $\delta$-Scuti variables based on observed light curves. The primary obstacle in this methodology is identifying an appropriate functional basis for the representation of the dynamical system. Indeed, this choice is frequently difficult in practice: in fact,  \citet{2022ApJ...930..161P} essentially failed at modelling RRab light curves using a polynomial basis. It follows that, at this stage, featurizing light curves based on the coefficients learned by SINDy seems unfeasible.

In this paper, we go beyond this limitation employing the Dynamic Mode Decomposition (DMD) algorithm, a mathematical tool introduced by \cite{schmid_2010}, and to the best of our knowledge, seldom used in astronomy before \citep[][]{2019MNRAS.490..114D,2020IAUS..353...65W, 2020AGUFMSH006..02L,2020arXiv200913095H, 2021RSPTA.37900181A,2021AGUFMSH53C..06L,2022ApJ...927..201A}. The advantage of this technique is that the extracted dynamic modes can be naturally interpreted as a generalization of global stability modes and can be potentially linked to the underlying physical mechanisms captured in the data sequence. In particular, we model the temporal evolution of the optical magnitude of a sample of RRLyrae stars belonging to the globular cluster $\omega$ Centauri. We automatically learn approximations for nonlinear dynamical systems encoding the time-evolution of the magnitude of each variable star and verify that the predicted evolution matches the observed data. Such approximations are expressed in terms of linear systems in suitable Koopman spaces, an eigenvector base of which is returned directly by the algorithm. The associated eigenvalues essentially summarize the light curve and can be in principle used as features for machine learning tasks. We find that the Koopman space required to describe RRc variable dynamics is smaller with respect to the one needed for encoding RRab variables dynamics. This reflects the simpler and more regular silhouettes of RRc curves and provides a straightforward heuristic for RRab vs RRc classification. Finally, we discuss an application to stars affected by the Blazhko effect \citep[long term variability; ][]{1907AN....175..325B}, which can be identified by comparing the eigenvalues corresponding to light curves measured at different stages in the Blazhko cycle. In one case we show that the Blazhko effect corresponds to a secular evolution in a small subset of eigenvalues only, a finding that may have physical significance.

\section{Photometric data}
For our work, we adopted the optical (BVI-band) photometry of the RRLs in 
the Globular cluster $\omega$ Centauri published by \citet{braga16, braga2019}. This data set is homogeneous in terms of photometric data reduction, even though it includes images collected during more than 20 years from seven different 1-to-8 m class telescopes, in a sky area of 57'$\times$56' (see Sec.2 in \citealt{braga16} for more details).
Still, to further improve the consistency of our analysis, we chose to adopt only the photometric data collected at the same telescope (Danish $1.5$m, at La Silla), during five years (1995 to 1999), which constitutes the largest part (almost $60\%$) of the original data set. The resulting Danish subset is a truly homogeneous data set, with the exception of sporadic positional offsets in the 1998 images.
As a result, the following analysis is based on light curves with $100$ to over $2000$ phase points in at least two of the three bands (V and B, since I-band phase points are one  order of magnitude less) for $125$ variables.

This subset of the data is comprised of light curves taken from $89$ RRc and $70$ RRab variables. Of these, $6$ RRc and $2$ RRab light curves were discarded because we failed to obtain an acceptable interpolation, due either to outliers or phase gaps. This leaves us with $83$ RRc and $68$ RRab light curves. Using the \emph{gcvspline} Python wrapper \citep[][]{lelosq} of a FORTRAN package for generalized, cross-validatory spline smoothing \citep[][]{woltring1986fortran}, we interpolate these light curves repeated $5$ times and picked equidistant (in phase) snapshots from this spline with $30$ points per phase which makes a smoothed out, periodic set of $150$ points from which DMD can build a model.
\section{Methods}
 Dynamic Mode Decomposition (DMD) algorithm is a dimensionality reduction algorithm guaranteed in the framework of Koopman analysis for the study of nonlinear dynamical systems.  The core principle of Koopman analysis is a non linear coordinate transformation that embeds a nonlinear dynamical system in an equivalent linear dynamical system of observables. The existence of such an embedding is due to an intrinsic trade-off between linearity and dimension. In this framework, it's necessary to figure out what the coordinate transformation is and what the dynamical system is turned into. For our purpose, we are interested to do this from data in one shot. This is possible thanks to tangent linear approximation of the system and a few arguments used in iterative methods for computing solutions to linear eigenvalue problems, that DMD algorithm naturally encodes.
 
More precisely, given a discrete time series $V_1^N=\{v_1,\dots,v_N\}$, where $v_i\in \mathbb{R}^m \hspace{0.2cm}\forall i \in [m]$, it is assumed the existence of a linear mapping $A\in GL_n$ that associates two successive points of the time series, i.e. $v_{i+1}=A v_i$. This last relation defines a linear dynamical system that stays approximately the same over the sampling period, formally using the language of matrices $V_2^N=AV_1^{N-1}+re_{N-1}^T$, where $r$ is the vector of residuals that encodes the behaviour not predicted by $A$, $e_{N-1}$ is the $(N-1)$th vector of the canonical base and $V_1^{N-1}=\{v_1,\dots,v_{N-1}\}$ and $V_2^{N}=\{v_2,\dots,v_N\}$. Roughly speaking, DMD computes the eigenvalues and the eigenvectors of $A$. Computing eigenvalues and eigenvectors is an elementary task for which several subroutines can be used; commonly, DMD uses Singular Value Decomposition (SVD) algorithm. Eigenvalues and the eigenvectors of $A$ completely define a set of modes each of which is associated with a fixed oscillation frequency and decay or growth rate. These modes are a natural generalization of the global stability modes and inherit their physical interpretation of encodings of damped or driven sinusoidal behavior in time. 

In this work, we deal with a low dimensional dynamical system, for this reason, we employ the extension of DMD known as \emph{Higher Order Dynamical Mode Decomposition} \citep[HODMD;][]{doi:10.1137/15M1054924}. The modification introduced by HODMD  is the following: instead of assuming $v_{i+1}=A v_i$, we consider an higher order Koopman assumption, i.e. that $v_{i+d}=A_1v_i+A_2v_{i+1}+\dots+A_dv_{i+d-1}$ where $d\geq 1$ is a tunable parameter. It is not difficult to notice that we are simply superimposing DMD on a sliding window over the time series. The main advantage of such an approach lies in enabling the computation of \emph{higher order modes} enriching the space returned by classical DMD when considering dimensional systems presenting low \emph{spatial complexity} but a large number of hidden frequencies.  

In our experiments, the time series $V_1^N=\{v_1,\dots,v_n\}$ is given by the uniformly sampled value of magnitude measured at regular intervals over the interpolated light curve as described above.  

\subsection{Implementation details}
We use the library PyDMD \citep[][]{demo2018pydmd}. We take $30$ snapshots per phase (for a total of $150$ snapshots) so that enough snapshots are fed to DMD to produce a model that fits with the original data but not too much as to model certain parts of the spline that may be due to noise. 
\begin{figure*}
    \centering
\includegraphics[scale=0.32]{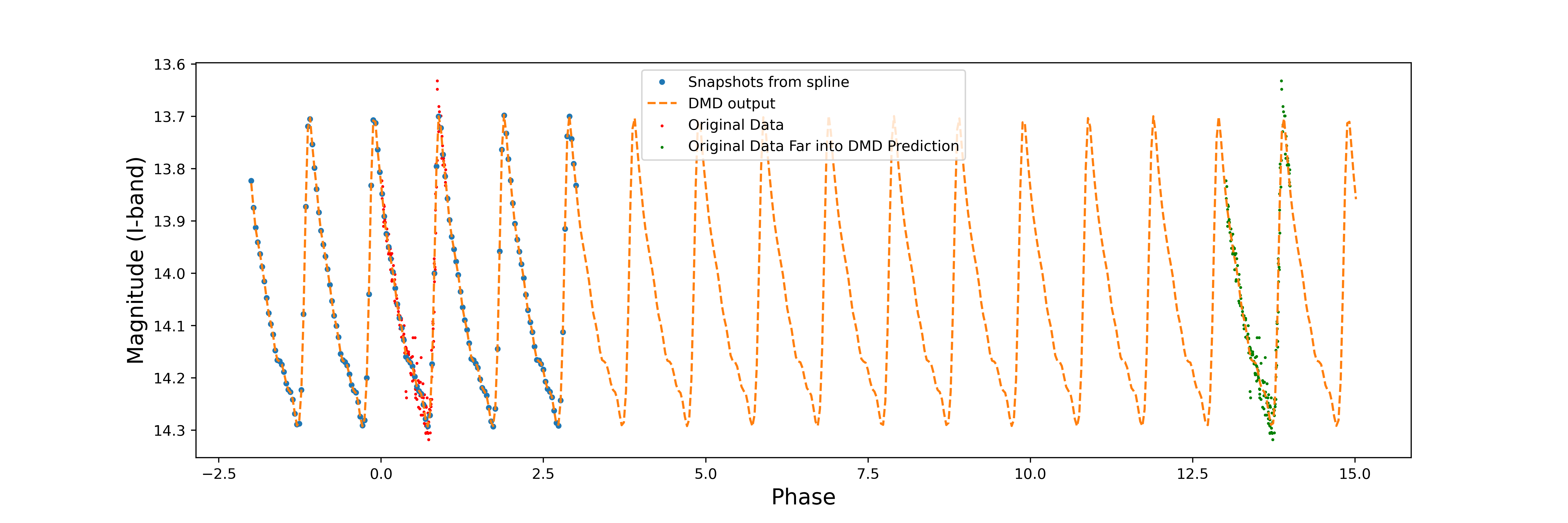}
    \caption{DMD reconstruction (orange dashed line) of the interpolated light curve (blue points) compared with the original (red points) and time shifted (green points) raw data for RRab star V$59$.}
    \label{fig:RRab_example_reconstruction}
\end{figure*}
\begin{figure*}
    \centering
    \includegraphics[scale=0.32]{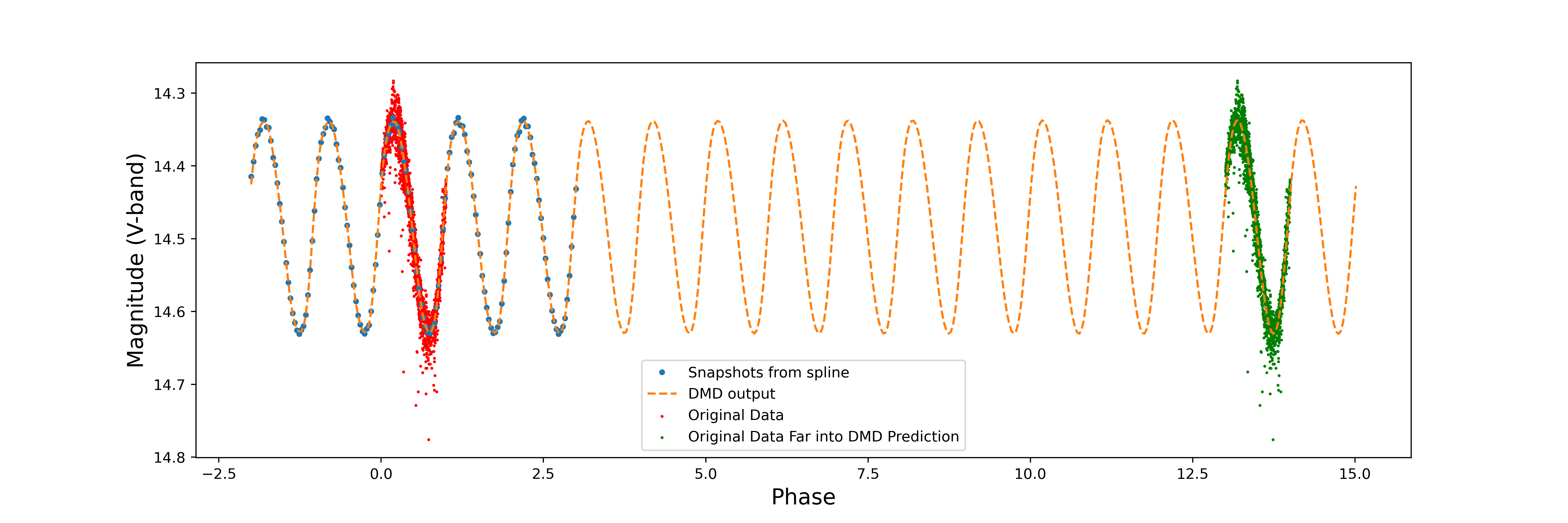}
    \caption{DMD reconstruction (orange dashed line) of the interpolated light curve (blue points) compared with the original (red points) and time shifted (green points) raw data for RRc star V$103$.}
    \label{fig:RRc_example_reconstruction}
\end{figure*}
\section{Results}\label{Results}
The DMD reconstruction of the light curve for an example RRab is shown in Fig.~\ref{fig:RRab_example_reconstruction} and for an RRc in Fig.~\ref{fig:RRc_example_reconstruction}.

\subsection{Separating RRab from RRc variables}
\begin{figure}[]
    \centering
    \begin{tabular}{c}
    \includegraphics[width=0.40\textwidth]{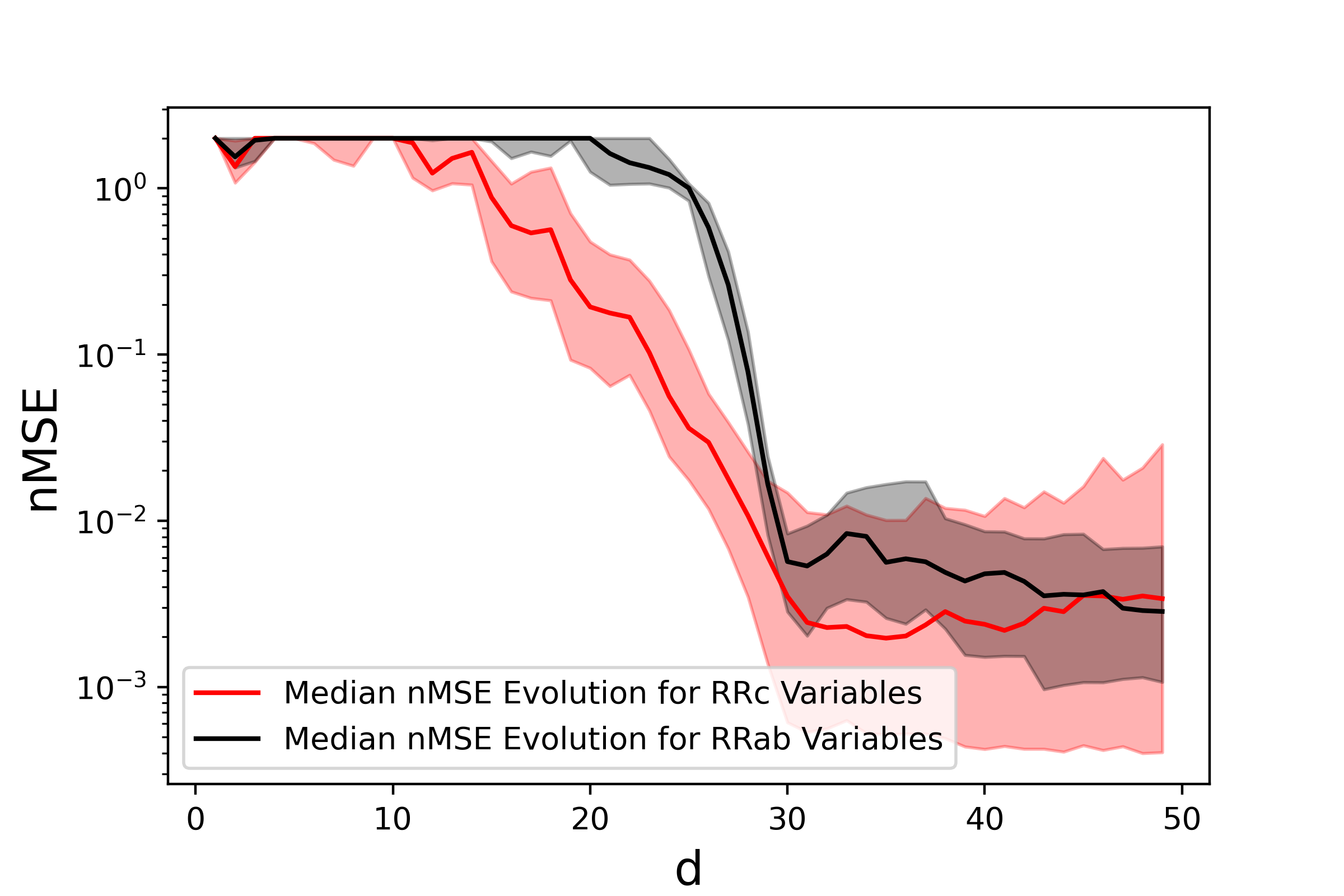} \\
    \includegraphics[width=0.40\textwidth]{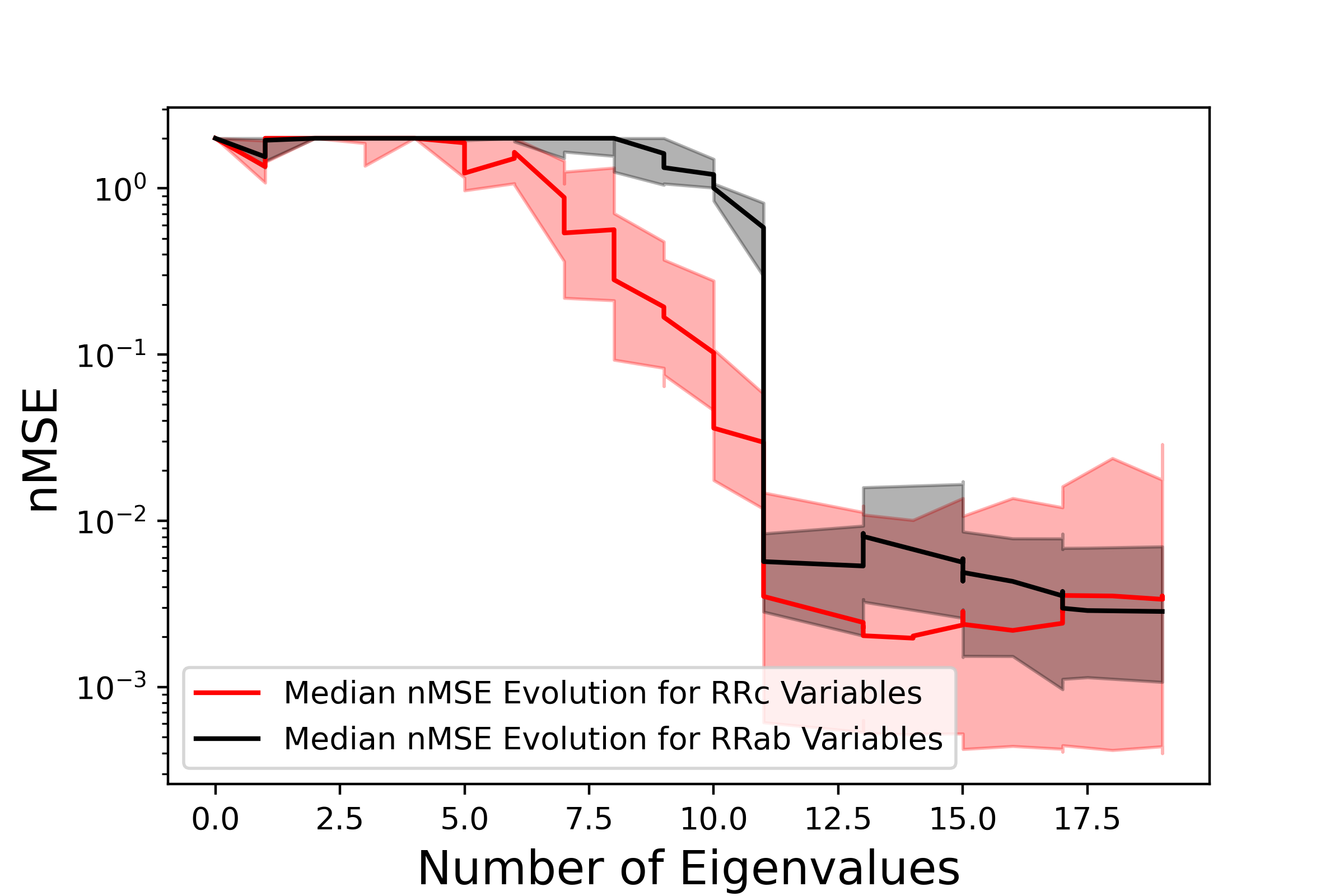} 
    \end{tabular}
    \caption{Left panel: nMSE achieved by the DMD reconstruction against the original light curve as a function of the spatial dimension of the HODMD snapshot ($d$ parameter in pyDMD). The black solid line represents the median curve for the RRab stars, the red solid line for the RRc stars. The lower and upper boundaries of the black and red shaded areas represent the first and third quartiles respectively. Right panel: Same as the left panel but nMSE is plotted as a function of the number of eigenvalues.}
    \label{fig:MSEdeig}
\end{figure}

As we vary the Koopman dimension $d$, which corresponds to the maximum allowable number of eigenvectors, we find that DMD models from RRc and RRab variables respond differently to increasing $d$. In fact, as the spatial dimension increases, the normalized mean squared error (nMSE; defined here as the ratio between the mean squared error (MSE) of the reconstructed light curve and the MSE of a constant fit) for models of RRc variables drops off steadily compared to models of RRab variables which have a more or less constant nMSE until about $d=25$ where we can observe a sharp drop. This is shown in Fig.~\ref{fig:MSEdeig} in the top panel and, for the number of eigenvalues, in the bottom panel. Note how nMSE starts decreasing immediately with $d$ for the RRc and keeps decreasing in a linear fashion, while it suddenly drops at about $d = 25$ for the RRab. From observing this phenomenon, we can choose a nMSE threshold (the nMSE is calculated $10$ phases away from the last snapshot to promote stability of the DMD model) such that, between RRc and RRab variables, different numbers of spatial dimensions (and different numbers of eigenvalues) are sufficient to produce models which yield an error equal or lower than the threshold.

\begin{figure}[H]
    \centering
    \begin{tabular}{c}
    \includegraphics[width=0.40\textwidth]{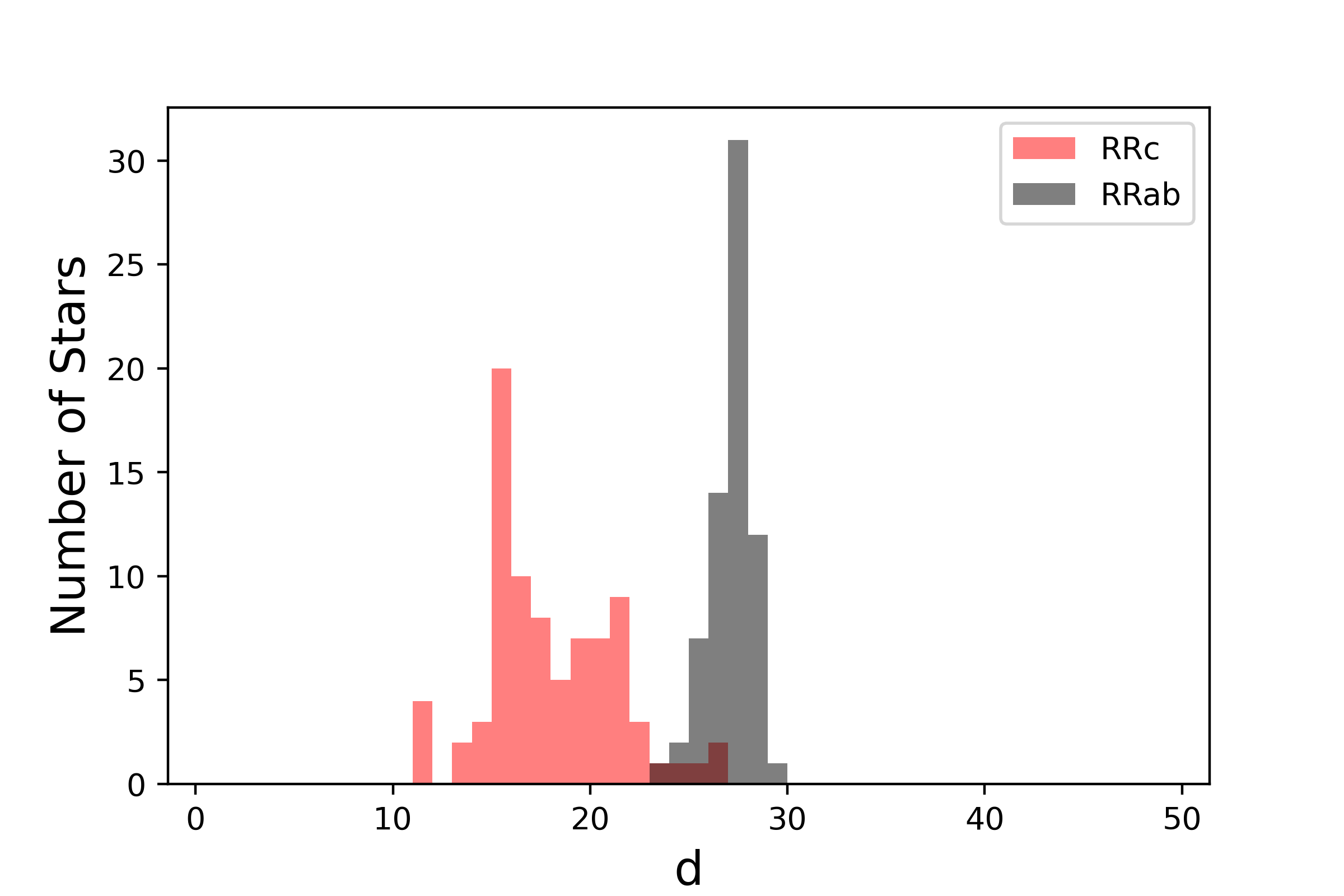} \\
    \includegraphics[width=0.40\textwidth]{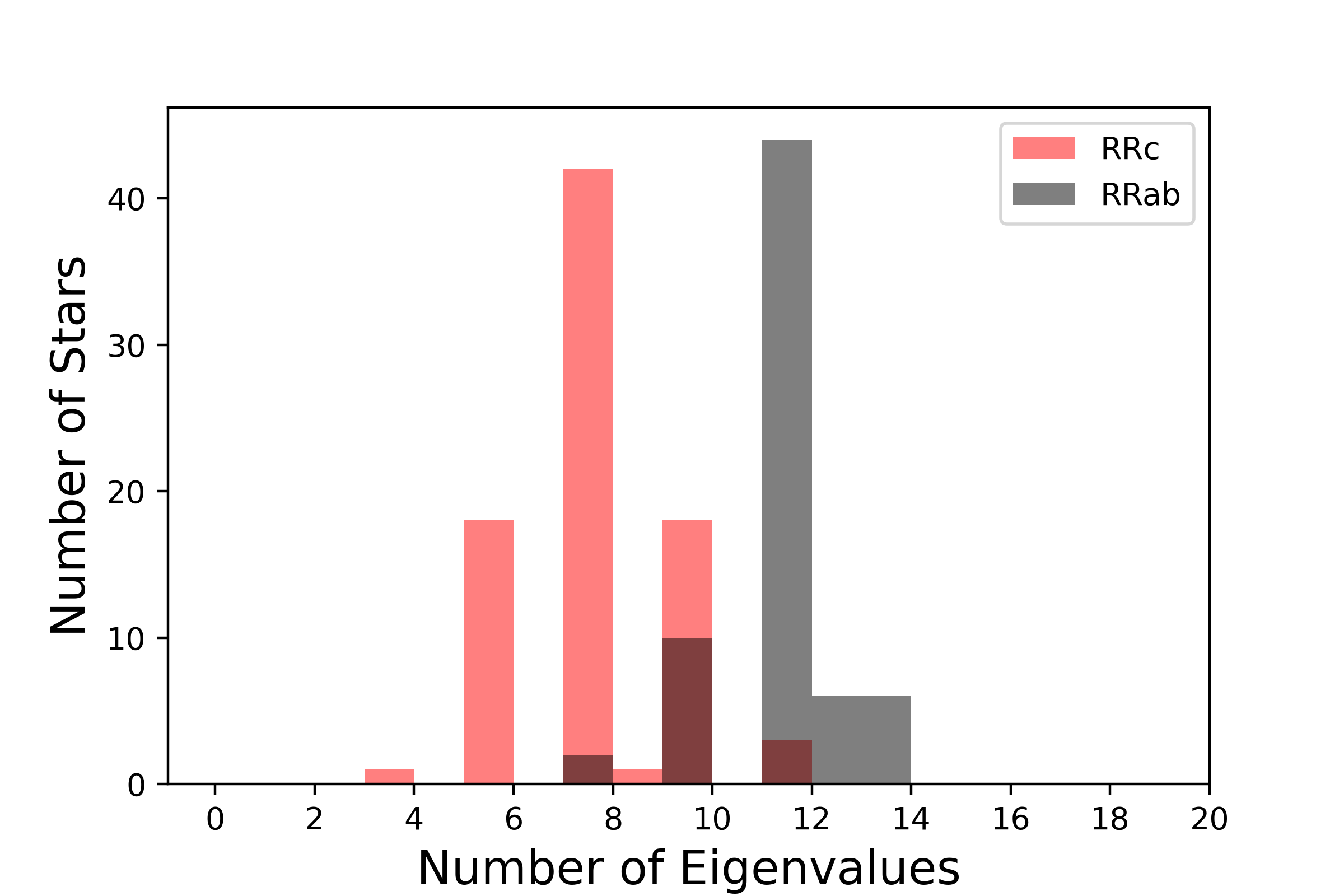} \\
    \end{tabular}
    \caption{Top panel: histogram of the minimum values of $d$ needed to achieve nMSE $=0.45$ (or below) for the RRab (black) and the RRc variables (red). Right panel: same, but for the number of eigenvalues.}
    \label{fig:d_hist}
\end{figure}

We find that, for the nMSE to be below $0.45$, models for RRc variables require between $11$ and $26$ spatial dimensions (d) and between $3$ and $11$ eigenvalues. For RRab variables, between $23$ and $29$ spatial dimensions (d) and between $7$ and $13$ eigenvalues are needed to yield an nMSE below threshold. We notice that there is some overlap especially in the numbers of eigenvalues. Despite that, this difference leads to a straightforward classification algorithm. In fact, we can find that, most likely, a variable star is RRc if $d<24$ is sufficient to yield an nMSE below $0.45$ and is RRab otherwise. The histogram of the values of $d$ needed to achieve nMSE$=0.45$ is shown in Fig.~\ref{fig:d_hist}. 

\begin{figure}[]
    \centering
    \includegraphics[width=0.40\textwidth]{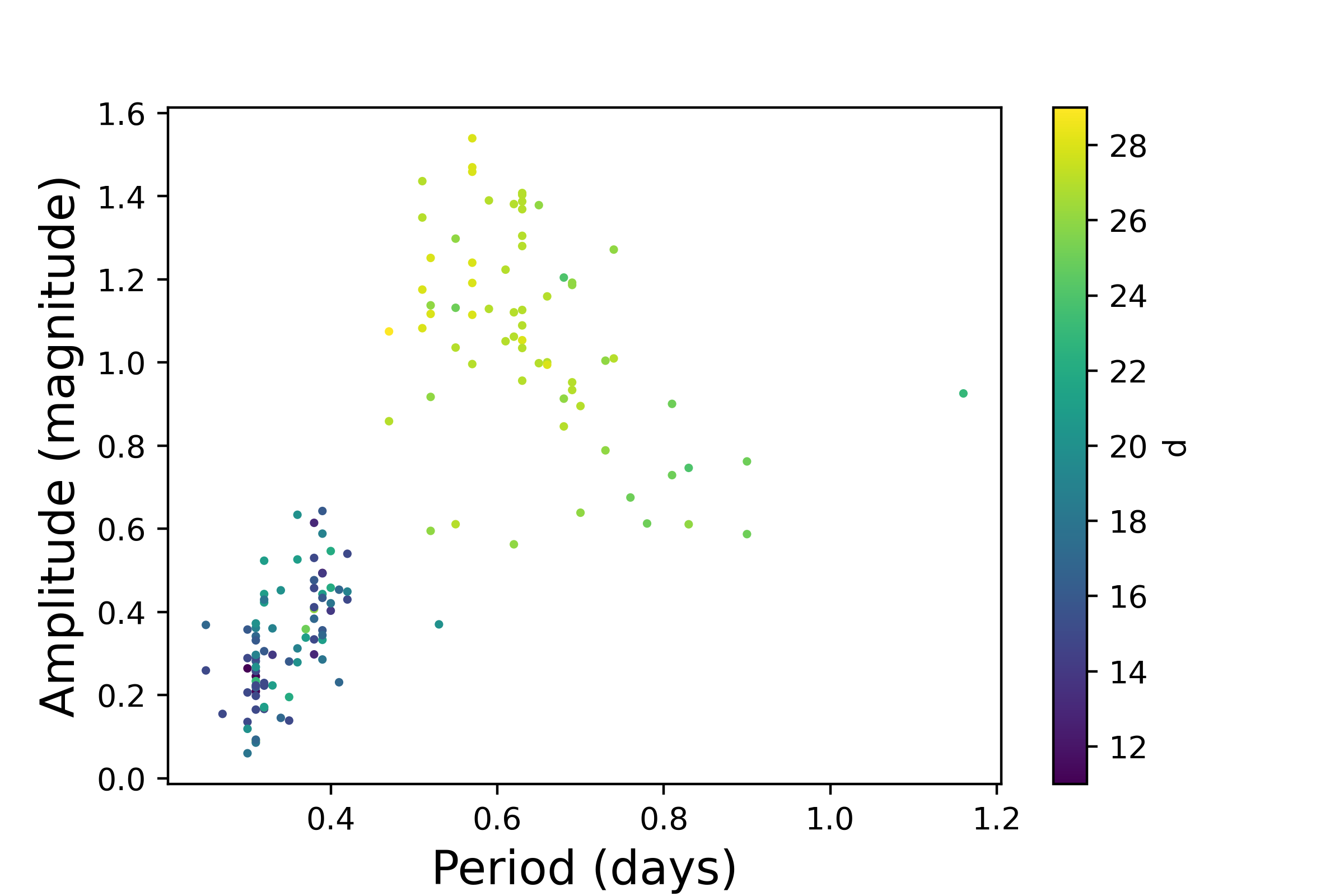}
    \caption{Bailey diagram with the Koopman dimension $d$ needed to attain nMSE $=0.45$ color coded.}
    \label{fig:dimension_bailey}
\end{figure}

The relevant confusion matrix for the associated classifier with $d<24$ is shown in Tab.~\ref{tab:confusion}. This is calculated on the same data set on which the threshold $d=24$ was chosen, so the associated measure of accuracy ($97\%$) is optimistically biased and should not be used to predict performance on unseen data. However this is still an indication that the modes identified by DMD are naturally suited to classification.
One more point to note is that classification based on the Koopman dimension $d$ needed to achieve a given level of mean squared error (nMSE=$0.45$), which depends on the shape of the light curve, overlaps quite well with the distinction between RRab and RRc variables in the Bailey diagram, as shown in Fig.~\ref{fig:dimension_bailey}. Discrepancies between the two could be leveraged to single out potentially interesting stars, among which perhaps mixed-mode oscillators. 

\begin{table}[]
    \centering
    \begin{tabular}{lcc}
    \hline
\hline
         & $d<24$ & $d\ge24$\\
         \hline
         RRab & 1 & 67\\
         RRc & 79 & 4\\
         \hline
    \end{tabular}
    \caption{Confusion matrix for a classifier based on the minimum Koopman dimension $d$ needed to achieve nMSE$=0.45$.}
    \label{tab:confusion}
\end{table} 

\begin{figure}[H]
    \centering
    \begin{tabular}{c}
    \includegraphics[width=0.30\textwidth]{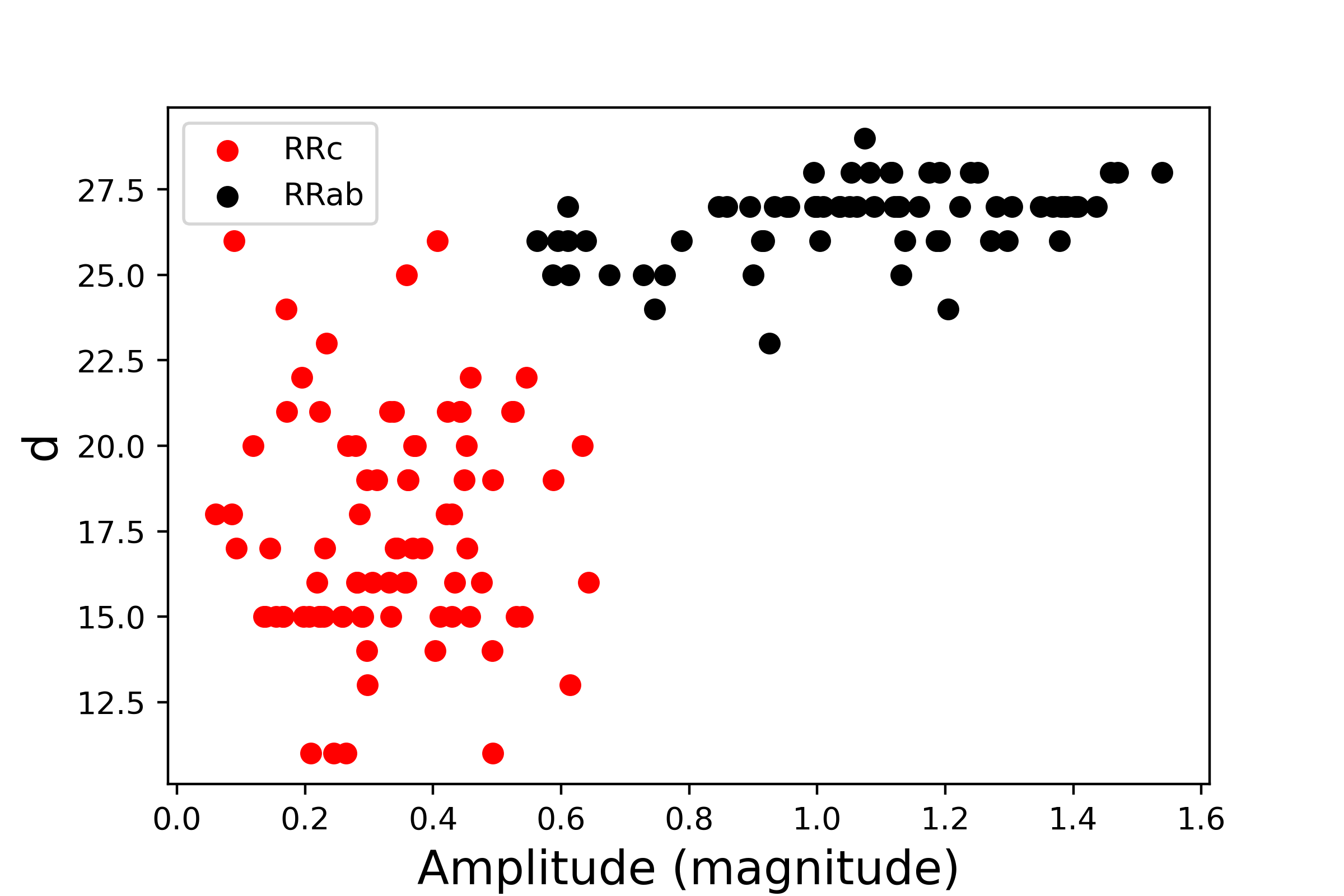} \\
    \includegraphics[width=0.30\textwidth]{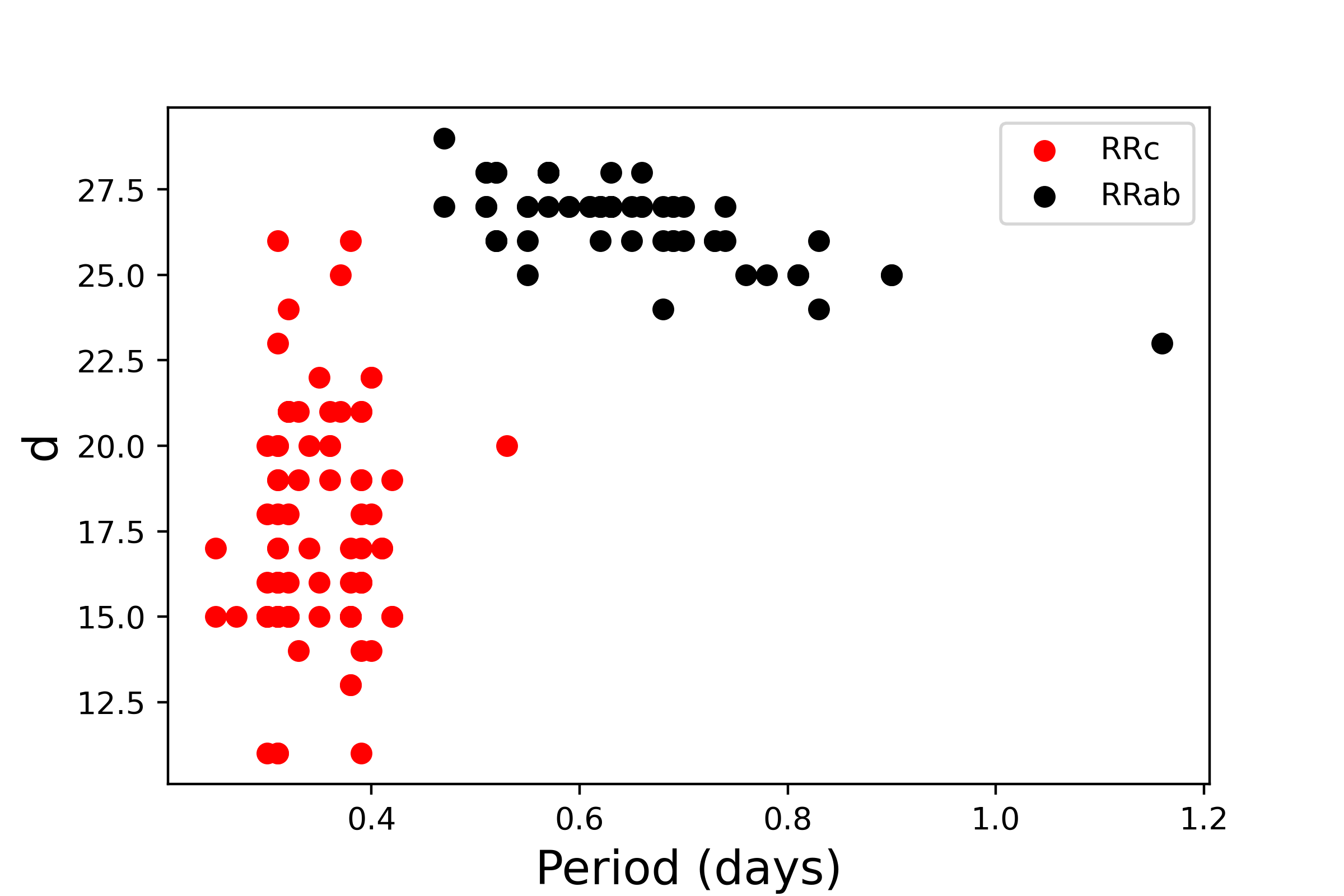} \\
    \end{tabular}
    \caption{Top panel: Koopman space dimension $d$ as a function of amplitude for RRab (black points) and RRc (red points). Right panel: Koopman space dimension $d$ as a function of period, same color coding as the top panel.}
    \label{fig:d_ampe}
\end{figure}

\begin{figure}[H]
    \centering
    \begin{tabular}{c}
    \includegraphics[width=0.30\textwidth]{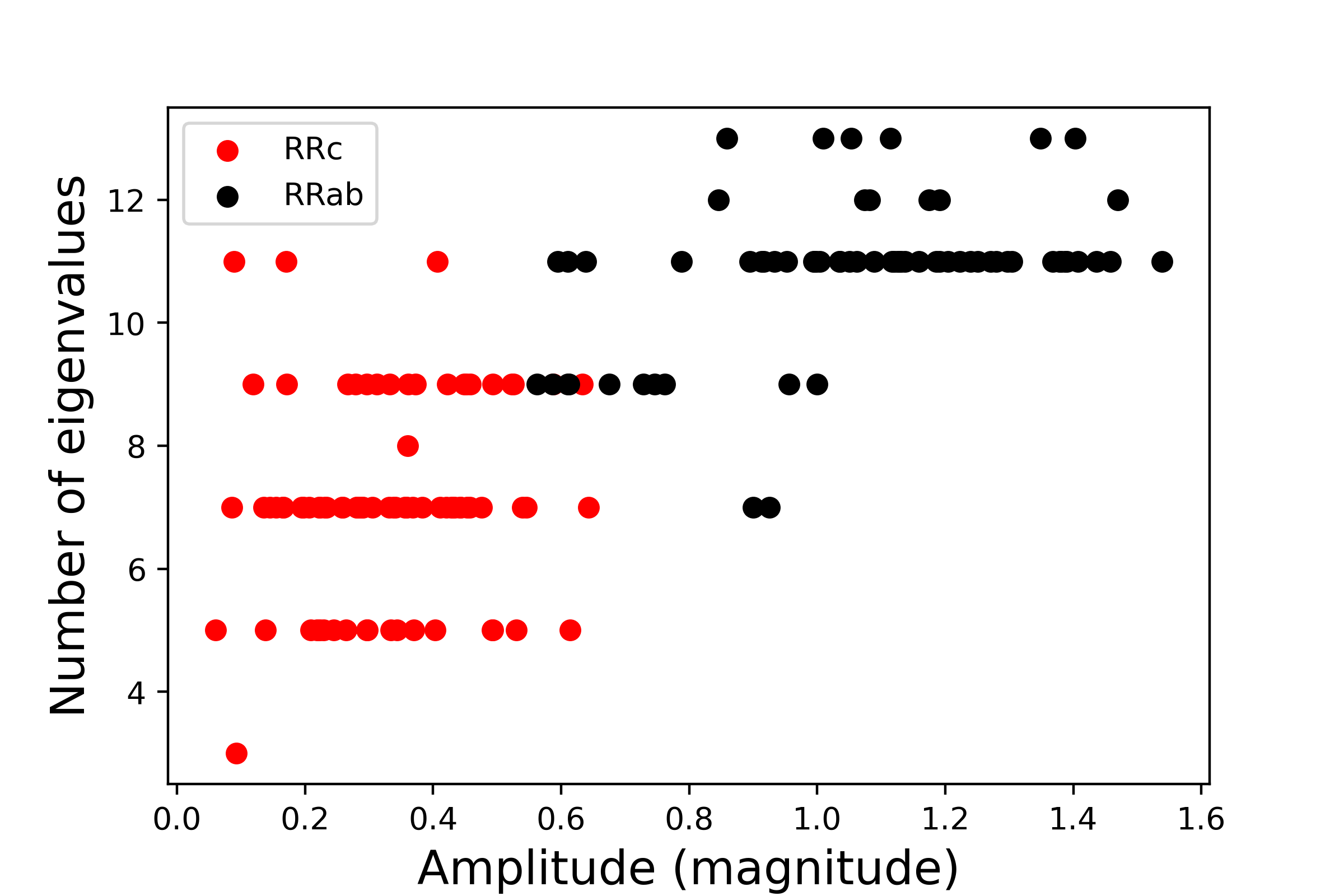}\\
    \includegraphics[width=0.30\textwidth]{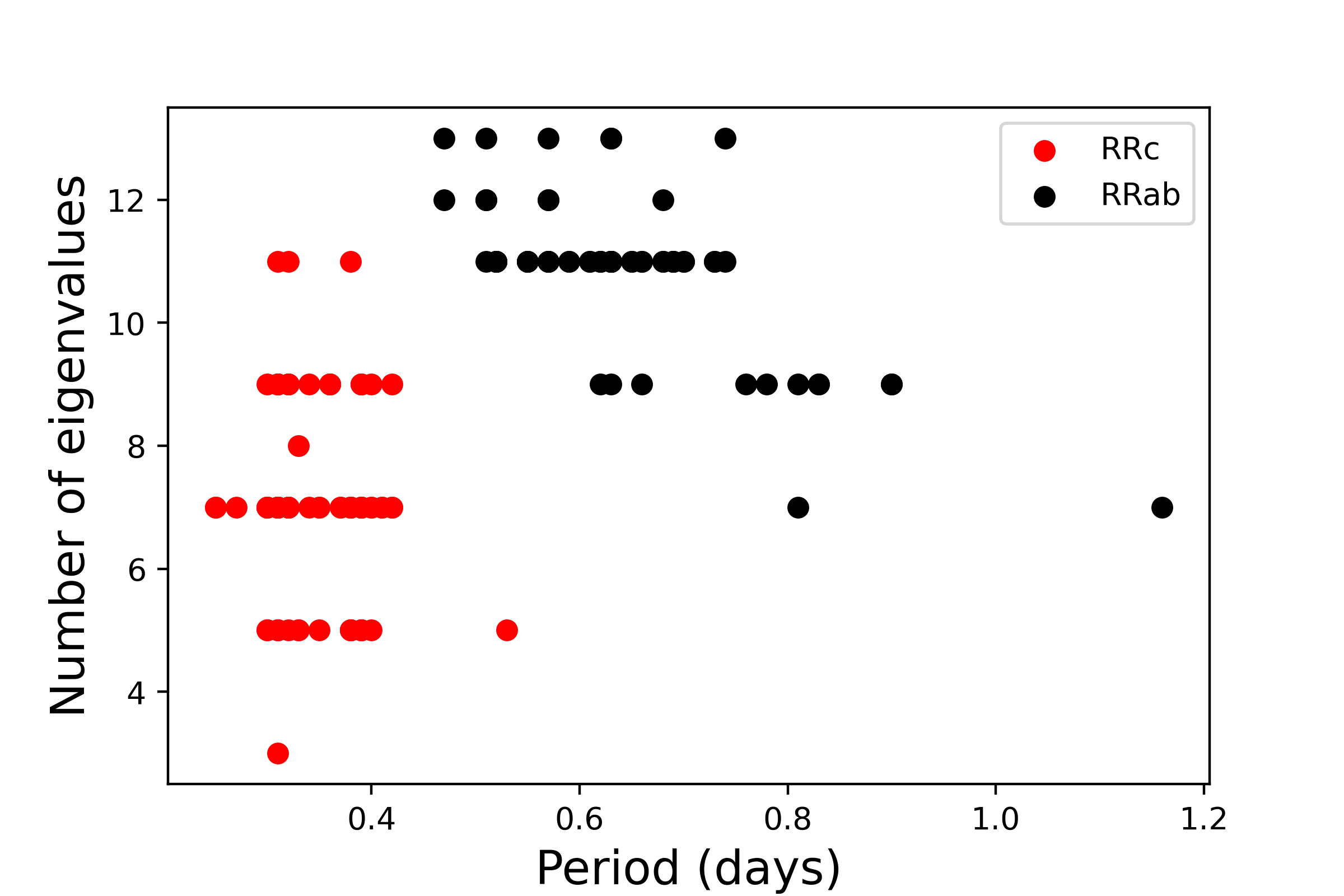} \\
    \end{tabular}
    \caption{Top panel: number of eigenvalues as a function of amplitude for RRab (black points) and RRc (red points). Right panel: number of eigenvalues as a function of period, same color coding as the top panel. It is known that short period, large amplitude RRab are harder to fit with fourier and in our case they also need more eigenvalues.}
    \label{fig:eig_ampe}
\end{figure}

A different way to look at the data presented in Fig.~\ref{fig:dimension_bailey} is shown in Fig.~\ref{fig:d_ampe}, where we plot $d$ as a function of the amplitude of the magnitude variation of each star (top panel) and as a function of the period (right panel). Additionally, in Fig.~\ref{fig:eig_ampe} we show the number of eigenvalues as a function of the same two quantities.
Regarding the dependence on amplitude, both $d$ and the number of eigenvalues increase with amplitude. Since both are measures of the complexity of the light curve shape, this is an expected result: higher amplitude oscillators become increasingly anharmonic. However, the increase is largely driven, in both cases, by the fact that large amplitude variables are RRab and small amplitude variables are RRc. Within each class the trend is weaker but still visible for the RRab, and absent for the RRc. Regarding period, the general trend is again increasing for both quantities, but it is decreasing for the RRab. This means that even though RRab stars typically have both a longer period and a higher number of eigenvalues than RRc, within the class of the RRab an increasing period corresponds to a \emph{lower} number of eigenvalues, i.e. a simpler light curve. 
\begin{table*}[H]
\centering
\begin{tabular}{llllll}
\hline
\hline
Class: &     & RRab & & RRc & \\
\hline
Parameter 1 & Parameter 2 & $R$ & $p$ value & $R$ & $p$ value\\
\hline
$N_\mathrm{eig}$&  Amplitude  & $0.45$ & $1.2 \times 10^{-4}$ & $0.13$ & $0.26$\\
$N_\mathrm{eig}$&  Period  & $-0.68$ & $2.3 \times 10^{-10}$ & $-0.08$ & $0.46$\\
$d$&  Amplitude  & $0.45$ & $1.0 \times 10^{-4}$ & $0.05$ & $0.68$\\
$d$&  Period  & $-0.71$ & $1.1 \times 10^{-11}$ & $0.09$ & $0.41$\\
\hline
\end{tabular}
\caption{Spearman correlation coefficients (col. 3, col. 5) between the number of eigenvalues and $d$ with amplitude and period over each class of variable stars and the associated p-values (col. 4, col. 6). \label{corre}}
\end{table*}

This evidence fully supports theoretical and empirical findings concerning the variation of RRab pulsation properties across the instability strip. The RRab display a steady decrease when moving from the blue to the red edge of the instability strip. The variation also applies to the shape of the light curves, indeed RRab close to the blue edge display a sawtooth shape and a more sinusoidal shape approaching the red edge. The main culprit causing this difference is convection, since it becomes more efficient when moving from hotter (bluer) to cooler (redder) effective temperatures and eventually quenches the pulsation activity \citep[][]{thisisamemsaiIbettherealarticleissomewhereelse}.

In terms of physical interpretation, it is particularly interesting that the MSE drops for RRab when $d$ approaches $30$, the number of points sampled along a period by our interpolation strategy. Since the data we fed to HODMD repeats exactly after one period, we have that $x_{31} = x_1$ and in general $x_{30 + k} = x_k$ for every $k$. Thus the vector to which DMD is effectively applied at $t+1$ is $\mathbf{x}_{t+1} = (x_2, x_3, ..., x_{30}, x_1)$ and at $t$ it is $\mathbf{x}_t = (x_1, x_2, ..., x_{29}, x_{30})$ and the matrix $\mathbf{A}$ such that $\mathbf{x}_{t+1} = \mathbf{A} \mathbf{x}_t$ becomes simply $[\mathbf{e}_{30}, \mathbf{e}_{1}, \cdot\cdot\cdot, \mathbf{e}_{29}]$. The eigenvalues of this matrix are $\lambda_k = e^{ki\pi/30}$ for $k = 0, ..., 29$, so they are equally spaced on the unit circle and include $\lambda_0 = 1$. We can thus conclude that the DMD is predicting the future evolution of the light curve by leveraging solely the periodicity of the data and nothing else; in an intuitive sense we can claim that an important fraction of the RRab curves is `incompressible'.

\subsection{Blazhko effect}
\begin{figure}[]
    \centering
    \begin{tabular}{c}
    \includegraphics[width=0.26\textwidth]{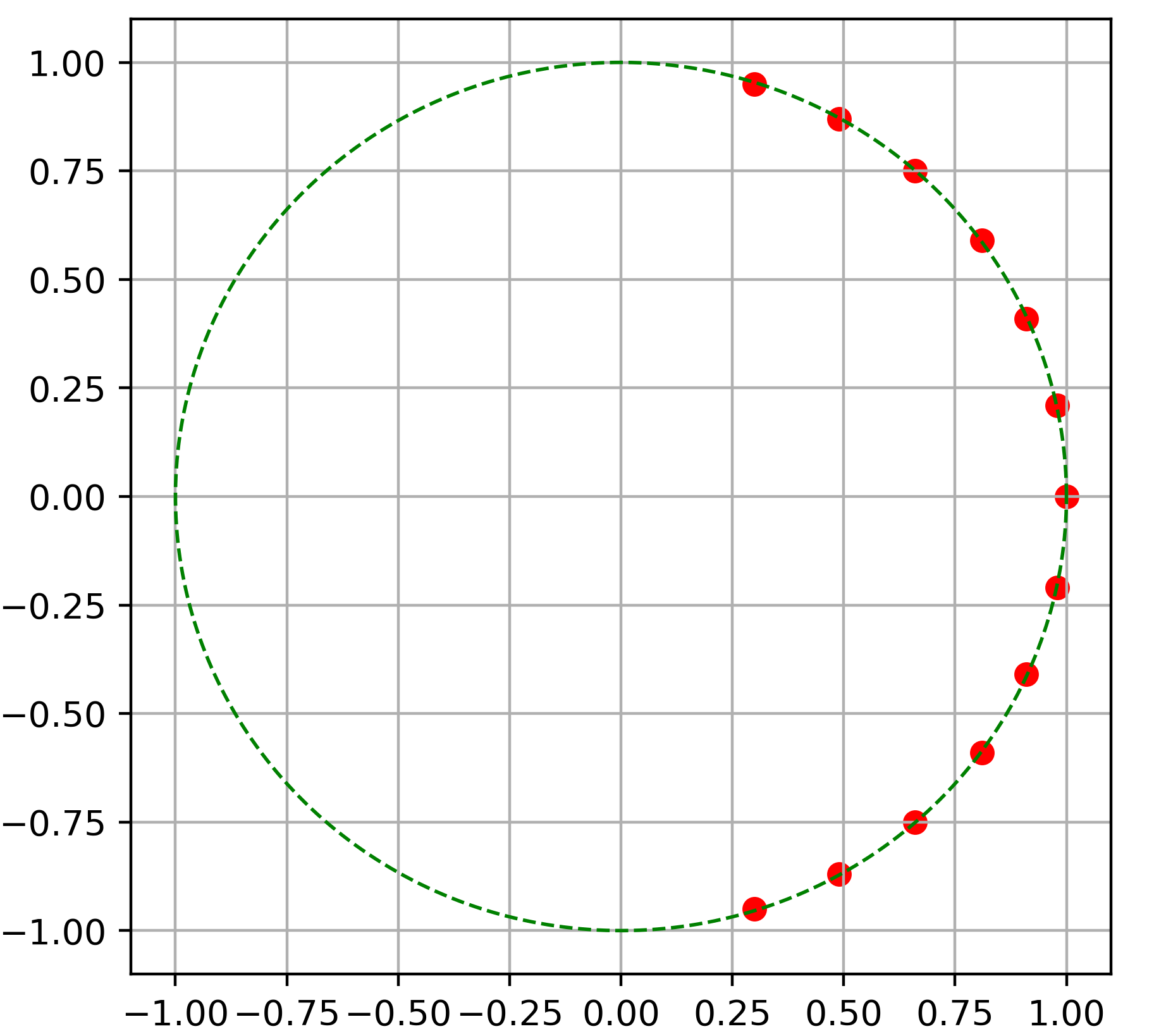}\\ \includegraphics[width=0.25\textwidth]{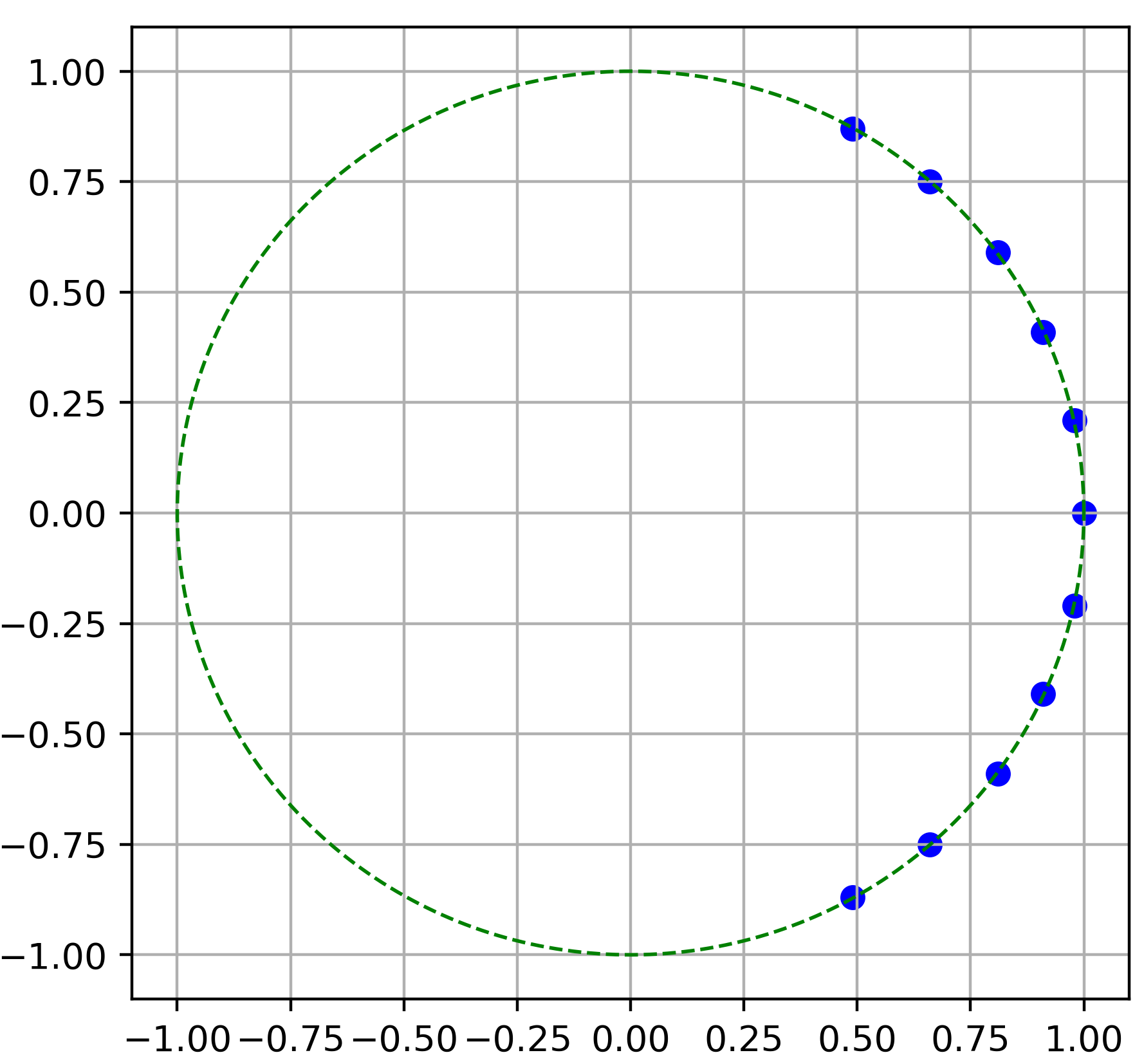}\\
    \includegraphics[width=0.25\textwidth]{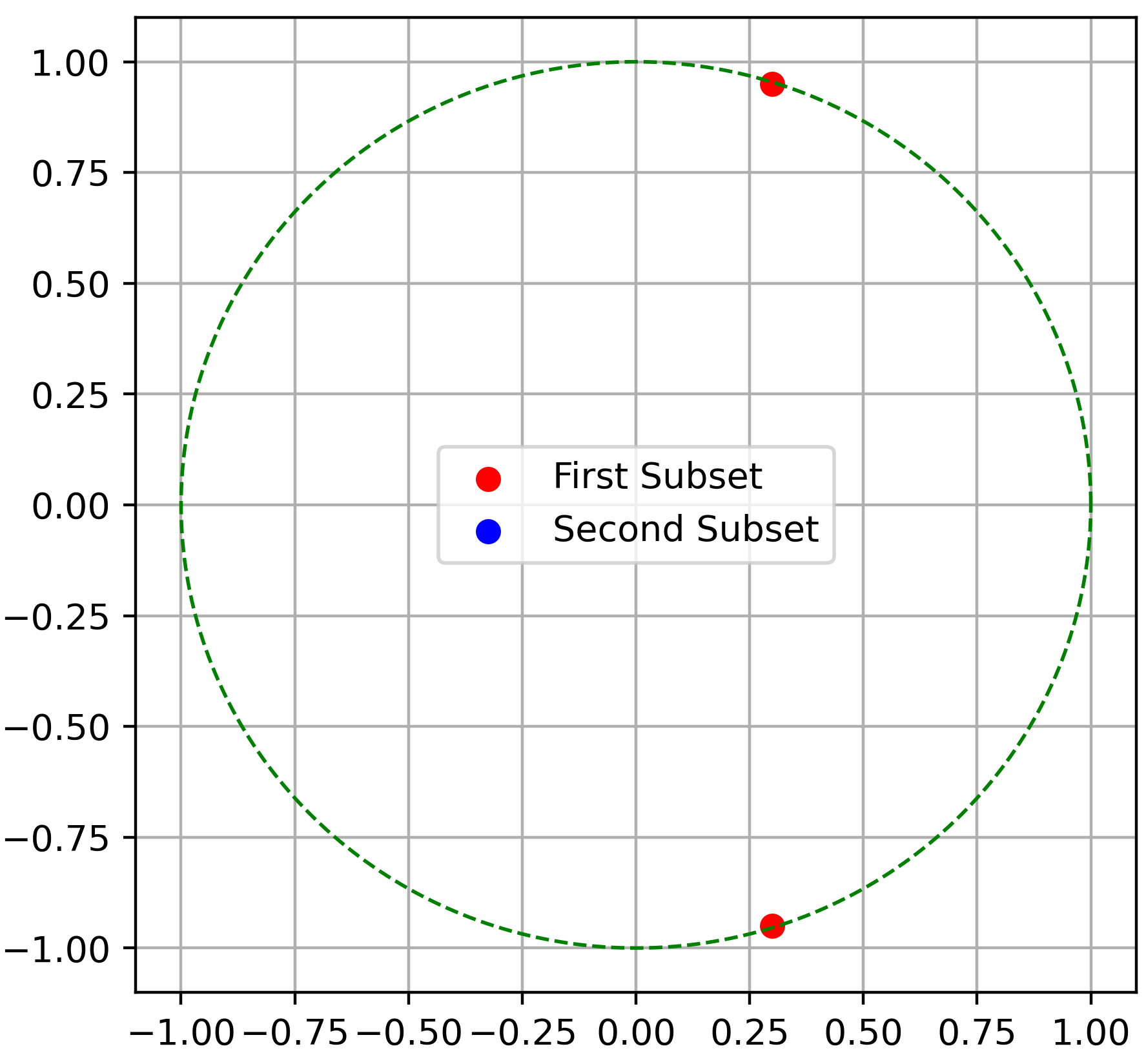}
    \end{tabular}
    \caption{Top panel: DMD eigenvalues (red points) for variable V$5$ calculated between Julian time $2449858.7698$ and $2449869.7601$, corresponding to the beginning of the observation period. The green dashed line corresponds to the unit circle in the complex plane. Middle panel: DMD eigenvalues (blue points) for variable V$5$ calculated between Julian time $2450971.5723$ and $2450985.6906$, corresponding to the end of the observation period. Right panel: DMD eigenvalues that differ between the two.}
    \label{fig:eigs12_V5}
\end{figure}

A final application of this description that goes beyond distinguishing fundamental and first overtone pulsators is related to Blazhko effect. DMD summarizes a light curve with a handful of complex numbers: do these differ over time for a Blazhko variable, and how?
In Fig.~\ref{fig:eigs12_V5} we show the DMD eigenvalues calculated during the first part of the observation period (top panel) and during the second part (right panel). Most eigenvalues match exactly except for a very limited number.

\begin{figure}{}
    \centering
    \begin{tabular}{c}
    \includegraphics[width=0.25\textwidth]{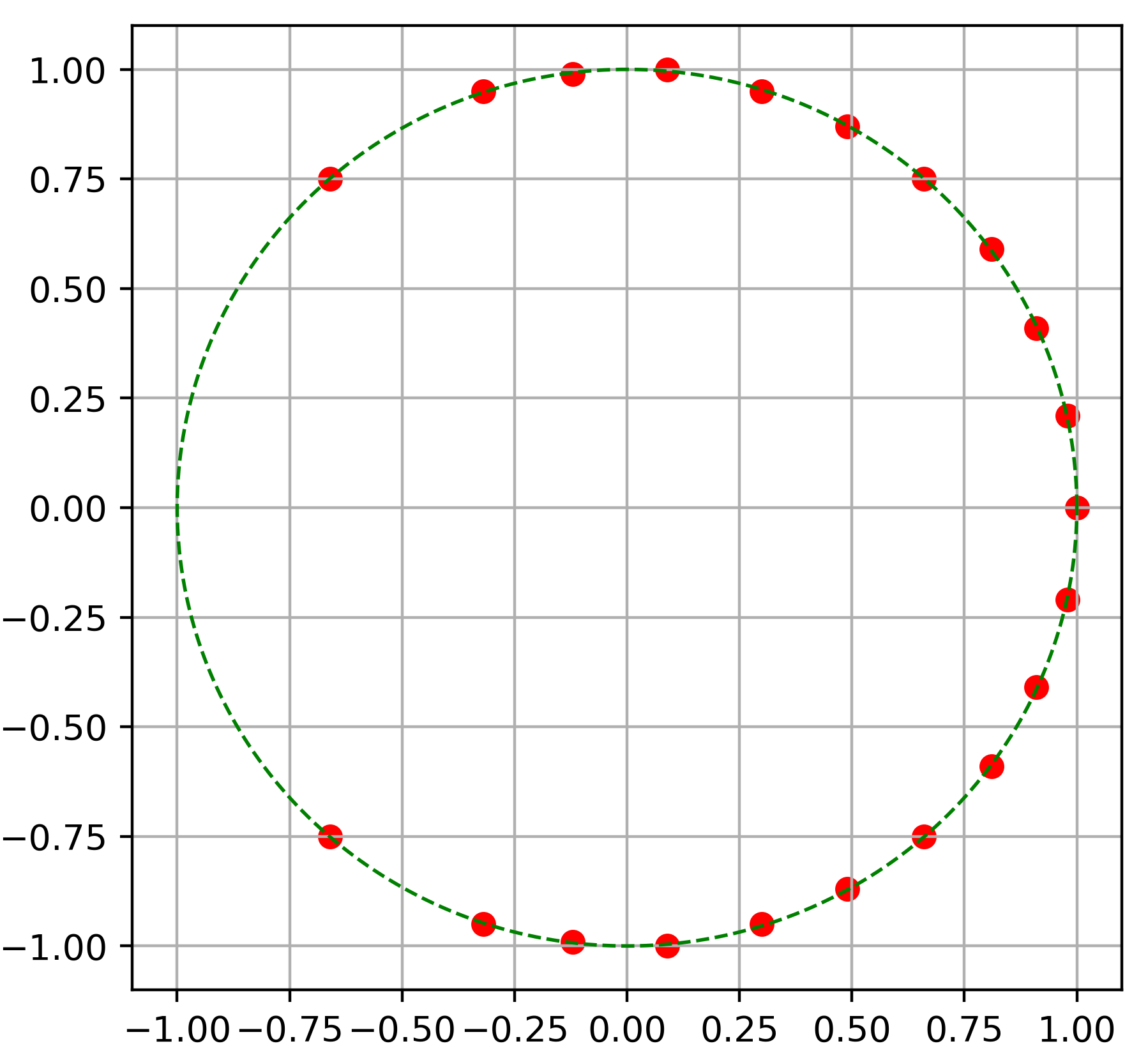} \\
    \includegraphics[width=0.25\textwidth]{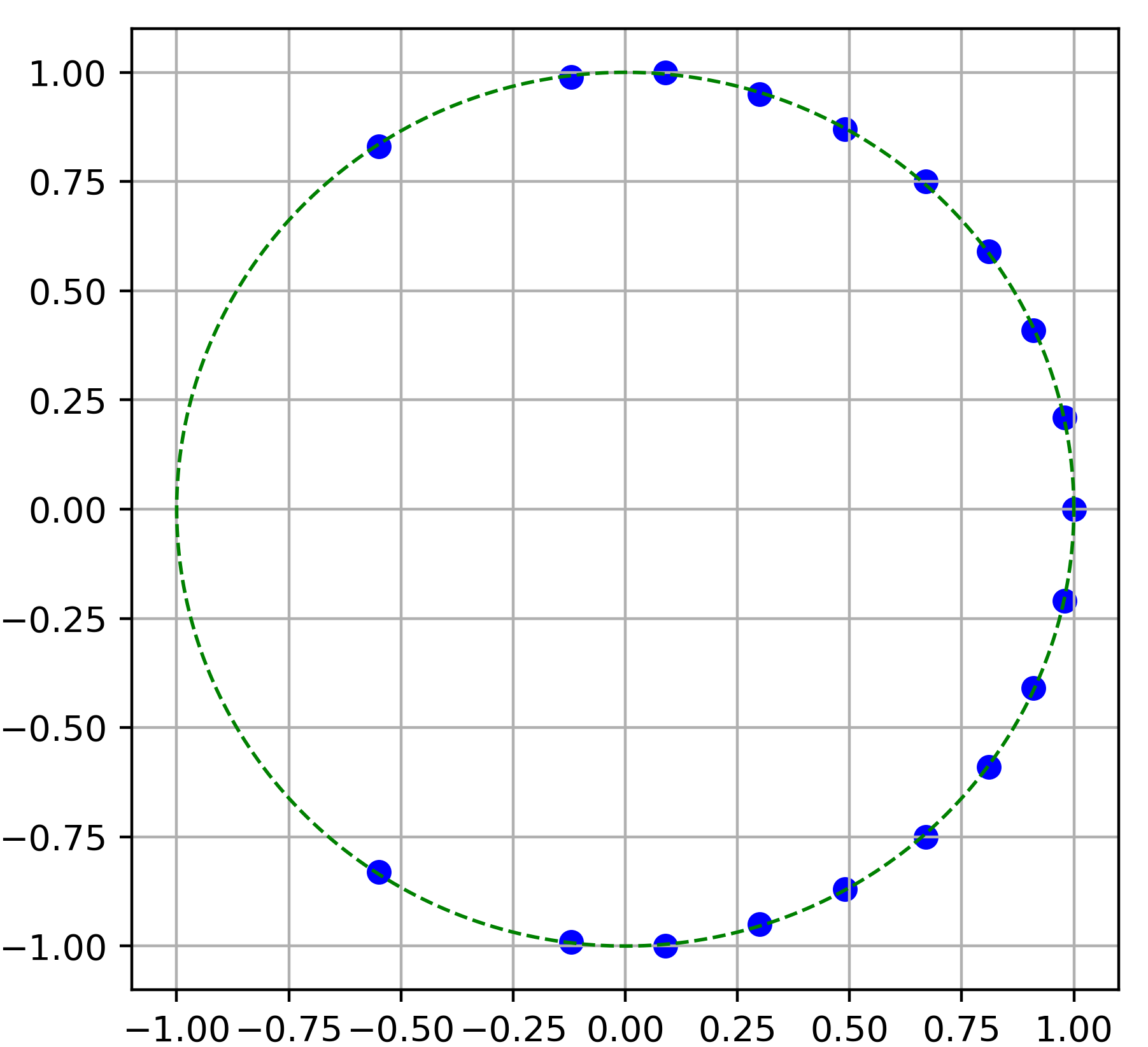}\\ \includegraphics[width=0.25\textwidth]{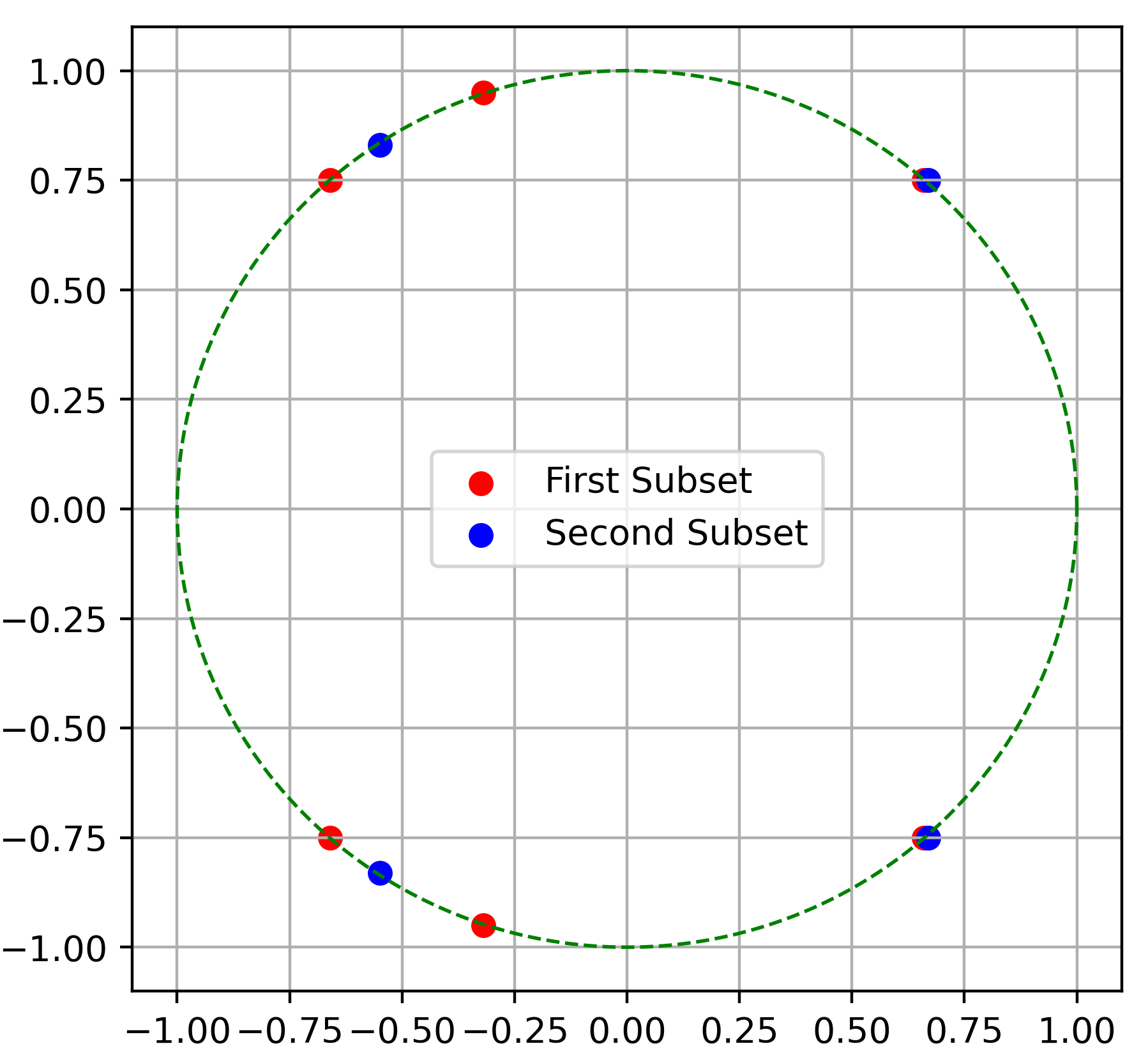}
    \end{tabular}
    \caption{Top panel: DMD eigenvalues (red points) for variable V$120$ calculated between Julian time $2449858.7468$ and $2449869.7635$, corresponding to the beginning of the observation period. The green dashed line corresponds to the unit circle in the complex plane. Middle panel: DMD eigenvalues (blue points) for variable V$120$ calculated between Julian time $2450970.5019$ and $2450985.6906$, corresponding to the end of the observation period. Right panel: DMD eigenvalues that differ between the two.}
    \label{fig:eigs12_V120}
\end{figure}

\begin{figure}
    \centering
    \begin{tabular}{c}
    \includegraphics[width=0.25\textwidth]{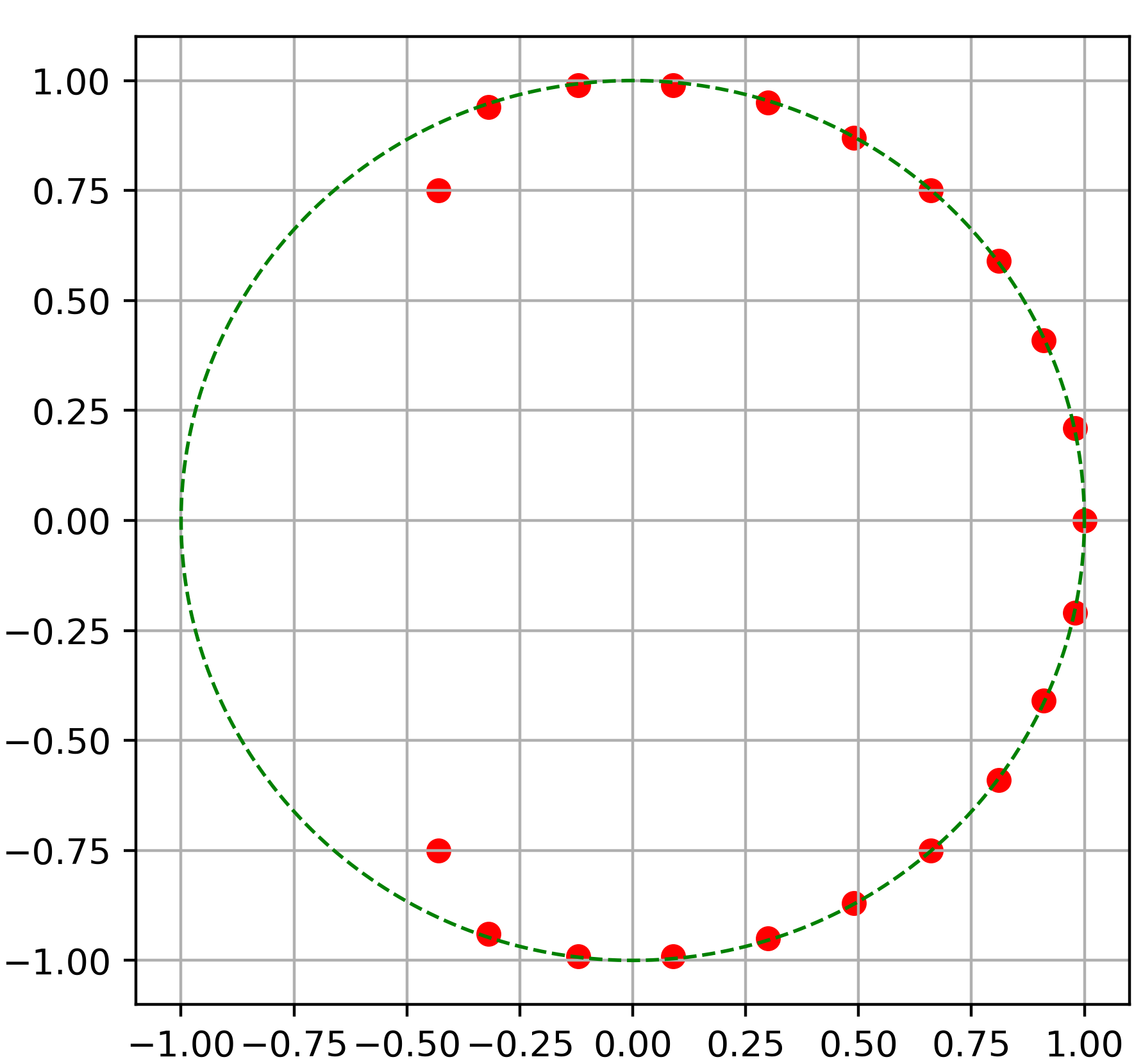}\\
    \includegraphics[width=0.25\textwidth]{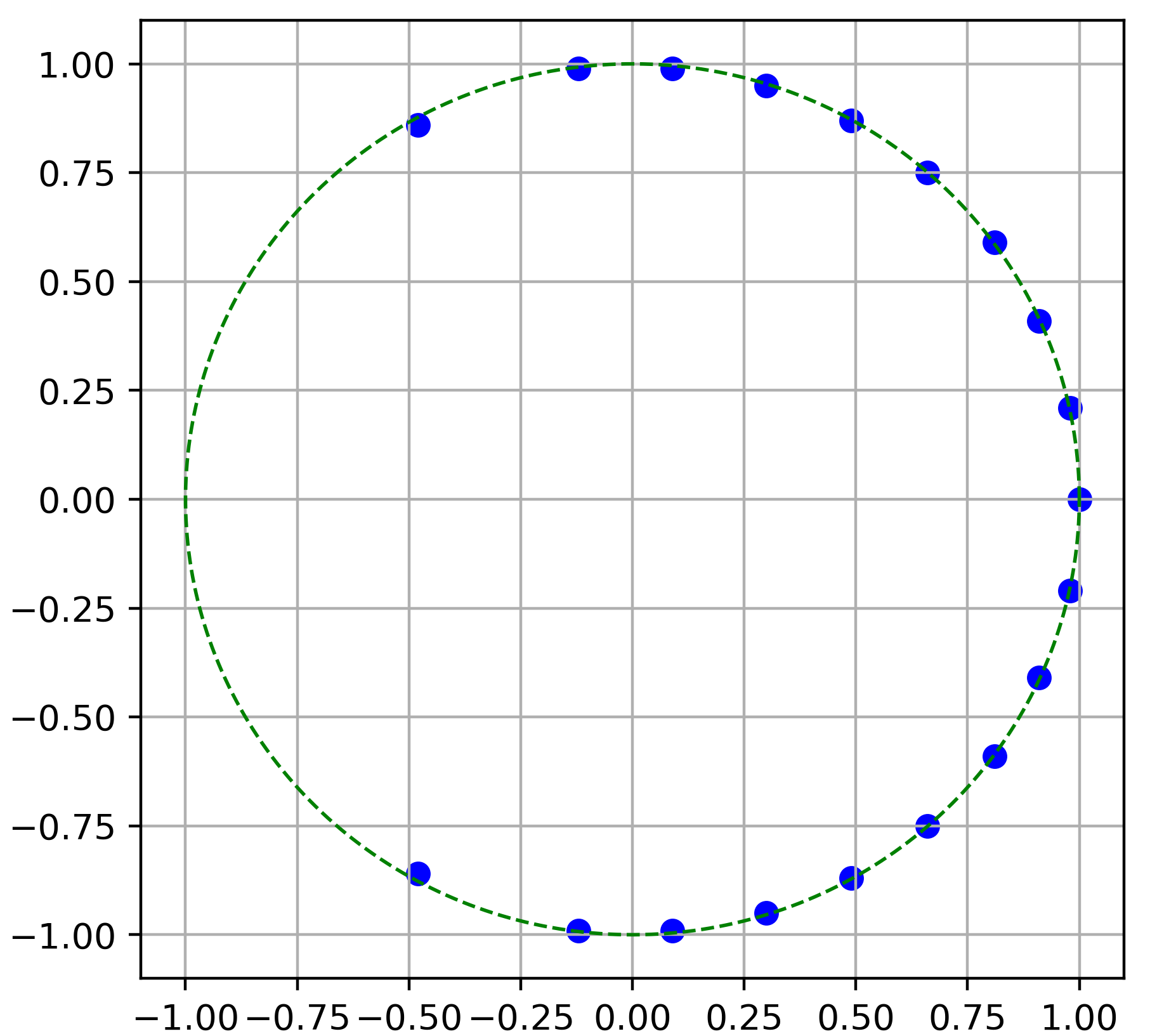}\\
    \includegraphics[width=0.25\textwidth]{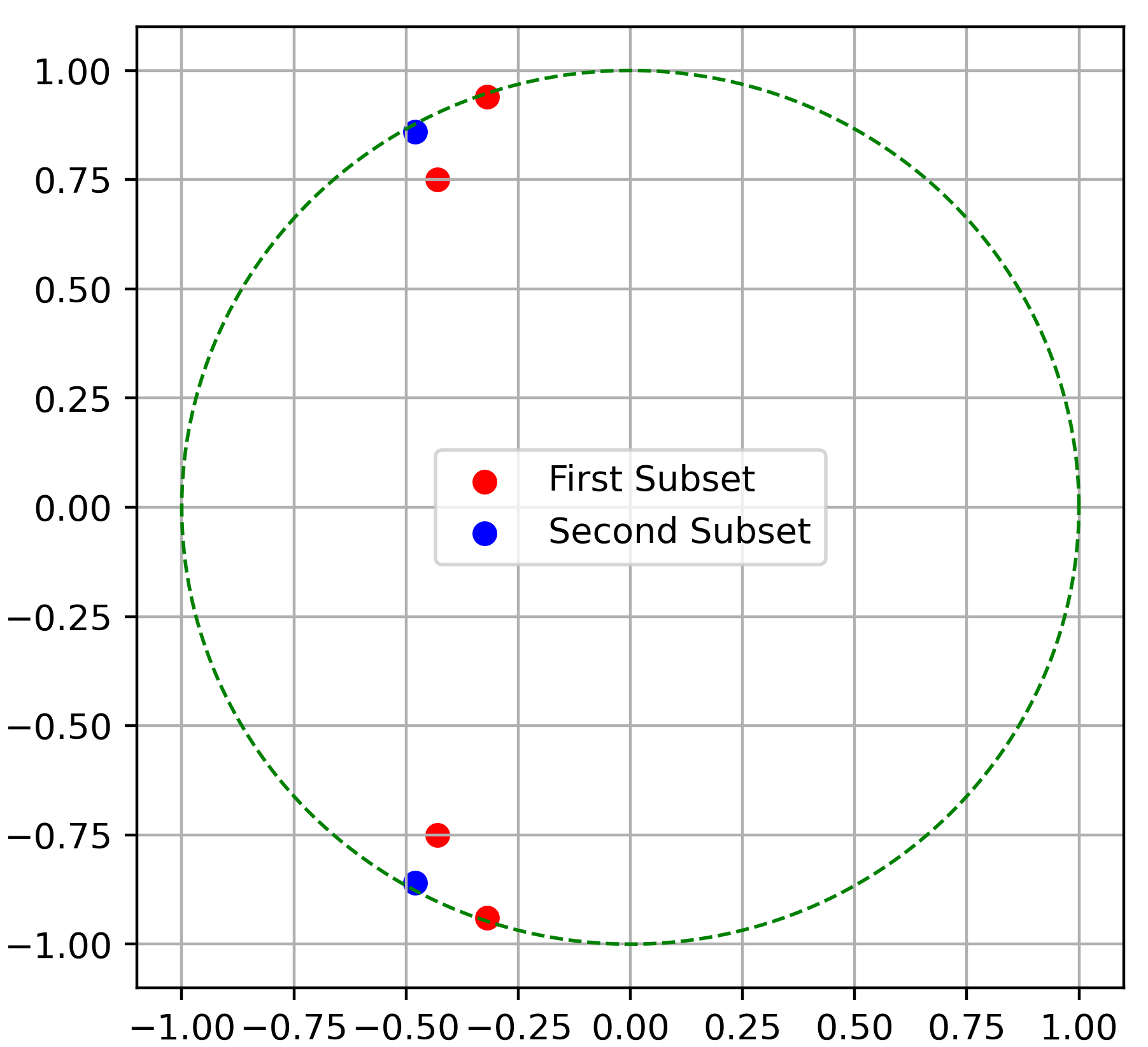}
    \end{tabular}
    \caption{Top panel: DMD eigenvalues (red points) for variable V$115$ calculated between Julian time $2449858.7468$ and $2449869.7635$, corresponding to the beginning of the observation period. The green dashed line corresponds to the unit circle in the complex plane. Middle panel: DMD eigenvalues (blue points) for variable V$115$ calculated between Julian time $2450971.5723$ and $2450985.6906$, corresponding to the end of the observation period. Right panel: DMD eigenvalues that differ between the two.}
    \label{fig:eigs12_V115}
\end{figure}

These eigenvalues may correspond to Koopman modes specifically associated to Blazhko evolution, since V$5$ is a suspected Blazhko. In any case DMD has revealed a secular change in its light curve. Potentially, this may be used to automatically detect Blazhko variables without the need for direct visual inspection of the light curve. For comparison, in Fig.~\ref{fig:eigs12_V115} and ~\ref{fig:eigs12_V120} we show the same for two more stars that are deemed affected by the Blazhko effect.


\vspace{-0.5cm}
\section{Conclusions}
 DMD and related algorithms have recently found application to fluid dynamics and other branches of physics that deal with high-dimensional snapshots either from observations or from simulations. In our application to variable stars, snapshots at a given time contain at most a few values, corresponding to magnitudes in various filters and possibly their derivatives. We thus applied the HODMD algorithm, which essentially synthesizes a higher-dimension snapshot from delayed coordinates. This is the first application of DMD-related methods to variable star light curves to date. This was done with the goal of obtaining features from light curves in a principled way that could be amenable to physical interpretation. In DMD the observed dynamics is decomposed into modes, similar to normal modes in a coupled oscillator system; each mode corresponds to an eigenvalue, associated either to exponential decay (or increase) or to periodic oscillations.

We obtained a reconstructed light curve as a linear combination of the individual evolution of such modes, and compared it with the original data over a sample of $151$ stars. We measured the discrepancy between the reconstructed light curve and the original by means of the mean squared error normalized to that of a constant fit. We are able to obtain consistently good fits (with nMSE under $10^{-2}$) if we let the number of eigenvalues increase as needed. We found that simpler light curves, such as those of RRc variables, require only a limited number of eigenvalues. Conversely RRab stars, which have more complex light curve shapes, require the inclusion of more eigenvalues to be described with good accuracy. We have quantified this, finding that that the Koopman dimension $d$ needed to achieve a given normalized MSE in reconstructing the curve behaves differently for RRab and RRc variables. We used this as a basis for a classifier, choosing a threshold of $0.45$ normalized MSE, to achieve which the modal RRab star requires $d = 27$ (resulting in a mode of $11$ eigenvalues) and the modal RRc star, $d = 15$ (corresponding to a mode of $7$ eigenvalues). 
The threshold value $0.45$ corresponds to the maximum (over all the light curves) of the minimum normalized MSE reached.

We found that the number of eigenvalues that needs to be included to obtain a good representation of the curve varies with amplitude and period in accord to the fact that RRab and RRc stars occupy distinctive loci in the period-amplitude plane. 

 While the DMD decomposition bears some resemblance to Fourier methods, which have been extensively applied to time domain astronomy and variable star light curves in particular, the advantage of DMD is to identify modes that are potentially physically meaningful. This is the case of the long term period variability due to the Blazhko effect, where indeed only a few eigenvalues found by DMD change between the beginning and the end of the observation period for the confirmed Blazhko variables in our data set. Whether this is indeed an indication of DMD finding specific modes associated to Blazhko evolution will be discussed in further detail in upcoming work. 

\section*{Acknowledgments}
This work is partially supported by Schmidt Futures, a philanthropic initiative founded by Eric and Wendy Schmidt as part of the Virtual Institute for Astrophysics (VIA).
M.P.\ acknowledges financial support from the European Union's Horizon 2020 research and innovation program under the Marie Sk\l{}odowska-Curie grant agreement No.\ 896248.
\newpage
\bibliography{example_paper}
\bibliographystyle{icml2024}

\newpage

\end{document}